\documentclass{llncs}

\usepackage{times}
\usepackage{graphicx}
\usepackage{amsfonts}
\usepackage{amsmath}
\usepackage{algorithm}
\usepackage{algpseudocode}
\usepackage{cite}
\usepackage{xspace}
\usepackage[pdfpagelabels=false]{hyperref}
\usepackage[caption=false,font=footnotesize]{subfig}
\usepackage{flushend}
\usepackage{epstopdf}
\usepackage{array}
\usepackage{multirow}
\usepackage{wasysym}

\newcommand{\sysname}{{CheapSMC}\xspace}

\DeclareMathOperator*{\argmin}{arg\,min}

\begin{document}

\title{\sysname: A Framework to Minimize SMC Cost in Cloud}

\author{Erman Pattuk\inst{1} \and Murat Kantarcioglu\inst{1} \and Huseyin Ulusoy\inst{1} \and Bradley Malin\inst{2}}
\authorrunning{Pattuk et al.}

\institute{The Univ. of Texas at Dallas \\
\email{erman.pattuk,muratk,huseyin.ulusoy@utdallas.edu}
\and
Vanderbilt University \\
\email{b.malin@vanderbilt.edu}}

\maketitle

\begin{abstract}
Secure multi-party computation (SMC) techniques are increasingly becoming more efficient and practical thanks to many recent novel improvements. The recent work have shown that different protocols that are implemented using different sharing mechanisms (e.g., boolean, arithmetic sharings, etc.) may have different computational and communication costs. Although there are some works that automatically mix protocols of different sharing schemes to fasten execution, none of them provide a generic optimization framework to find the cheapest mixed-protocol SMC execution for cloud deployment.

In this work, we propose a generic SMC optimization framework \sysname that can use any mixed-protocol SMC circuit evaluation tool as a black-box to find the cheapest SMC cloud deployment option. To find the cheapest SMC protocol, \sysname runs one time benchmarks for the target cloud service and gathers performance statistics for basic circuit components. Using these performance statistics, optimization layer of \sysname runs multiple heuristics to find the cheapest mix-protocol circuit evaluation. 
Later on, the optimized circuit is passed to a mix-protocol SMC tool for actual executable generation. Our empirical results gathered by running different cases studies show that significant cost savings could be achieved using our optimization framework.
\end{abstract}

\section{Introduction}
\label{sec:intro}

Over the last couple of years, many two party secure multi-party computation (SMC)  protocols have been  proposed to address different secure computation needs ranging from privacy-preserving face recognition (e.g.,~\cite{cite:sadeghi2010efficient}) to secure biometric identification (e.g.,~\cite{cite:evans2011efficient}). 
In addition, many generic two party circuit evaluation platforms (e.g.~\cite{cite:henecka2010tasty, cite:bogdanov2008sharemind}) have been developed to improve the efficiency of the existing secure protocols.  Most of these platforms (e.g.,~\cite{cite:malkhi2004fairplay}) also provide  high level programming language support that can automatically generate circuits from programs written in C like languages. These recent advances made two party SMC protocols ever more practical and created the possibility of wide spread deployment in practice. 

Still, there are other problems that need to be solved to make these platforms truly practical. One important challenge that received little attention is the performance optimization. Recent work \cite{cite:demmler2015aby} showed that different two party SMC protocols may have different computational and communication cost profiles. For example,  arithmetic sharing based circuit evaluation protocols may be better for certain tasks compared to Yao's garbled circuit evaluation protocols. On the other hand, boolean secret sharing based circuit evaluation techniques could be best in some situations. It has been shown~\cite{cite:demmler2015aby} that  combination of these three different techniques could perform much better than using any single one of them.  This raises the following research question: \textit{how to find the best combination of the different two party SMC techniques for a given task?}

Unfortunately, most of the existing work do not consider the problem of finding the optimal combination of different sharing based protocols for a given task and require the user to manually choose the specific sub-protocols. Although there were some recent work that tries to optimize and automate the selection process (e.g.,~\cite{cite:kerschbaum2014automatic}), that work had limited scope with respect to the cost dimensions that the user can choose to optimize. For example, if one party leverages a cloud infrastructure for running the protocol, the network communication may significantly impact the overall monetary cost paid by the parties. Therefore, we need an optimization framework that can automatically consider communication, computation and monetary costs when finding the optimal two party SMC protocol composition.

Our goal in this work is to build an optimization framework where the given two party SMC task could be automatically optimized under the given cost constraints; and the optimal (or near optimal) combination of different sharing based sub-protocols (e.g., arithmetic, boolean, Yao's secret sharing protocols, etc.) are selected automatically.  Our goal could be seen similar to other automatic task optimization frameworks seen in other systems. In our optimization framework, we especially focus on the cloud setting since the cloud computing is being widely adopted by organizations~\cite{cite:forbes} due to its flexibility and low initial cost of ownership. In the cloud setting, in addition to minimizing the overall run time of the system, we may need to balance the network traffic, and computation time to achieve overall lowest monetary cost. This makes the problem even more challenging since we may need to consider both communication and computation costs in the optimal mix-protocol circuit generation. 

\textbf{Overview of the \sysname.} The aim of our system is to make it easier for the parties to implement and execute SMC protocols, while minimizing the monetary cost of the SMC execution in the cloud. In order to ease the implementation phase and make our system easily extensible to available SMC tools, we divided \sysname into three main parts. The \emph{Programming API} acts as the front-end for the users that allows them to implement their SMC protocol using our C++ library. In the end, this layer is responsible for representing the user protocol as a circuit of atomic operations. Note that our system can be further extended by designing a custom language (e.g., SFDL of \emph{Fairplay}~\cite{cite:malkhi2004fairplay}), which uses our \emph{Programming API} in the background. In this work, we do not
focus on such integration and instead focus on the optimization aspects.

The \emph{Optimization Module} is responsible for assigning secret sharing schemes (e.g., Arithmetic, Boolean, Yao sharing as in the \emph{ABY} framework~\cite{cite:demmler2015aby}; Additively homomorphic, Yao sharing as in the \emph{Tasty} framework~\cite{cite:henecka2010tasty}, etc.) to the nodes in the circuit, such that the total cost of executing this protocol in cloud is minimized. Since finding an optimal solution to this optimization problem is NP-Hard, the optimization module provides heuristics to find near optimal solutions.

Finally, the \emph{SMC Layer} actually implements the \emph{optimized} circuit given some existing SMC tool (e.g., \emph{ABY}, \emph{Sharemind}~\cite{cite:bogdanov2008sharemind}). Since the proposal for new and more efficient SMC tools is a popular research topic, the design of \sysname does not focus on a single SMC tool that could limit its usefulness. Instead, we leverage a given SMC tool as a black-box, provided that it is a mixed-protocol SMC tool (i.e., allows implementation using different sharing schemes). 

\sysname proposes several atomic operations (e.g., addition, multiplication, binary xor, binary and, etc.) that covers various application scenarios. In Section~\ref{sec:case}, we show the results of applying our system to several different case studies, while further applications can simply be done using our C++ library. Moreover, the process of optimizing the circuit is totally decoupled from the circuit generation and other layers, so that proposing a new heuristic and actually implementing it can be achieved with minimal effort.

\textbf{Our Contributions.} They can be summarized as follows:
\begin{itemize}
	\item We formally define the problem of minimizing the monetary cost of running SMC applications in the cloud setting given for circuit based two party SMC protocols. We particularly focus on the monetary cost of running virtual machines (VMs) and transferring data over the network in the optimization objectives. Our profiling results in Section~\ref{sec:cost} show that the network transfer cost may significantly impact the overall cost. Therefore, making protocol selections based on just optimizing the running time may not be ideal (i.e., optimizing performance only).
	\item We propose an easily extensible system, \sysname, which uses existing SMC tools as a black-box, and propose two different novel heuristics in addition two existing baseline methods to address the optimization problem. Our heuristics specially designed for the circuit based SMC protocols and can be used with any circuit based SMC tool.
	\item We profile the cost of \sysname using Amazon EC2 cloud service. We investigate two different scenarios (i.e., \emph{Inter-Region} and \emph{Intra-Region} VM placement) and four different VM models. For each combination, we derive the average cost of executing atomic operations that highly depends on the VM model and the scenario.
	\item We apply our system to two applications (e.g., biometric matching, matrix multiplication) and evaluate the monetary cost reductions achieved by our platform. We compare our heuristics with existing optimization heuristic of Kerschbaum et al.~\cite{cite:kerschbaum2014automatic}, and pure garbled circuit~\cite{cite:yao1982protocols}. For the Inter-Region scenario, our heuristics gives up to $96\%$ and $30\%$ cheaper results compared to pure garbled circuit execution and Kerschbaum et al., respectively. 
\end{itemize}

\section{Background}
\label{sec:background}

\emph{Secret Sharing} is a well-known cryptographic primitive that allows a party to partition its secret input and share those partitions (called a \emph{share}) with other parties.~\cite{cite:shamir1979share}. Due to the nature of the secret sharing, none of the parties (except the party that owns the secret input) can learn anything about the secret input by just looking at a given share. A number of shares should be combined to \emph{reconstruct} the partitioned secret. 

In the remaining of the paper, we will heavily use several different two party secret sharing schemes, such as \emph{Arithmetic}, \emph{Boolean}, \emph{Yao}. We will briefly discuss them here.

\paragraph{Arithmetic Secret Sharing} In this secret sharing scheme, the secret is assumed to an integer $x$ in a ring $\mathbb{Z}_n$. The first party, who owns the secret $x$, generates a random number $x_1 \in \mathbb{Z}_n$ and subtracts it from the secret to get $x_2 \equiv x - x_1~\mod{n}$. $x_1$ is the first party's share, while $x_2$ sent to the second party is her share. This type of secret sharing is also called the \emph{Additive Secret Sharing}~\cite{cite:cramer2000general}. 

\paragraph{Boolean Secret Sharing} The key idea of this secret sharing scheme is similar to the previous one,  except the granularity of the sharing. In order to share a secret using \emph{Boolean} sharing, first the secret should be represented in its binary form, then each bit is shared separately in mod $2$. Given the Boolean shares of a secret, the parties can execute any operation using the Goldreich-Micali-Widgerson (GMW) protocol~\cite{cite:goldwasser1987play}. 

\paragraph{Yao Secret Sharing} It is also called the Yao's \emph{Garbled Circuit} protocol, and allows parties to execute any boolean circuit over any number of secret inputs~\cite{cite:yao1986generate}. The key idea of the garbled circuit protocol is that one party, the \emph{garbler}, randomly generates two encryption keys $k^w_0, k^w_1 \in \in \{0,1\}^K$ for each wire $w$ of the circuit. She then encrypts all combinations of the outputs using the generated keys and sends the encryption keys for the corresponding input wires of the party, along with the garbled circuit. The other party, the \emph{evaluator}, decrypts the wires using the keys at hand, and recurses through the garbled circuit by using the decrypted keys. An interested reader is referred to the original paper for the details~\cite{cite:yao1986generate}. 

\paragraph{Conversion Between Different Schemes} Since most SMC tools provide multiple sharing schemes in their implementation, they also provide conversion protocols from one scheme to another. However the certain steps of the conversion protocol may vary in different SMC tools. The most recent SMC tool, the \emph{ABY} Framework, allows conversion amongst Arithmetic, Boolean, and Yao sharing schemes~\cite{cite:demmler2015aby}. The \emph{Sharemind} protocol shows a different method on converting shares between Arithmetic and Yao sharing~\cite{cite:bogdanov2008sharemind}. \emph{Tasty} framework allows conversion between additively homomorphic sharing and Yao sharing~\cite{cite:henecka2010tasty} An interested reader is referred to those works for conversion protocols. 

\section{The Optimal Partitioning Problem}
\label{sec:problem}

\subsection{Problem Definition}
\label{sec:problem:definition}

Let $\mathcal{S}=\{s_1, \ldots, s_n\}$ be the set of provided secret sharing mechanisms. Then given a variable $x$ in some domain $\mathcal{I}$, let $[x]_{s_i}$ represent the secret sharing of $x$ using $s_i$. 

Next, we define the set of operations $\mathcal{O}=\{o_1, \ldots, o_k\}$ such that each operation $o_i\in\mathcal{O}$ takes a set of parameters that are secretly shared in $s_j$ and outputs a single variable secretly shared in $s_j$. Note that the number of inputs that an operation takes is fixed, regardless of the secret sharing scheme. An operation $o_i$ is \textbf{supported} in $s_j$, if there exists an execution protocol that takes the input parameters to $o_i$ and outputs a result secretly shared in $s_j$. Let $\delta(o_i)\subseteq \mathcal{S}$ represent the secret sharing mechanisms that support operation $o_i\in\mathcal{O}$. 

Since an operation can be executed in the cloud environment, one should approximate the monetary cost of performing the protocol execution in a certain setup. In order to achieve this, we focus on the processing and network transfer costs of executing a single operation in the pay-as-you-go cloud model. In this computing model, a customer of a cloud provider service is charged a constant amount per unit time for using a particular type of virtual machine (VM), while the prices vary depending on the processing capabilities of the VM. On the other hand, the monetary cost of transferring a single byte in and out of the VM is fixed based on the VM specifications. The unit cost of network transfer vary as the network capacity of the VM changes.

Under such circumstances, we define the processing and network transfer costs of executing an operation $o_i\in\mathcal{O}$ in the secret sharing scheme $s_j\in\delta(o_i)$ as $\mathtt{P}(o_i, s_j)$ and $\mathtt{N}(o_i, s_j)$, respectively. Furthermore, we define the processing and network transfer cost of converting a variable that is secretly shared in $s_i\in\mathcal{S}$ to $s_j\in\mathcal{S}$ as $\mathtt{CP}(s_i, s_j)$ and $\mathtt{CN}(s_i, s_j)$, respectively. Note that defined costs may vary based on VM specifications. 

Given the set of operations $\mathcal{O}$, the parties in the computation (i.e., the server and the client) implement a circuit $\mathcal{C}$ that is represented as a directed acyclic graph (DAG) and consists of $m$ nodes $c_1, \ldots, c_m$. Each node represents a single operation and takes input from other nodes, whereas the number of inputs is decided by the operation. To be more concrete, let $\alpha(c_i)\in\mathcal{O}$ represent the operation that is assigned to the node $c_i\in\mathcal{C}$, while $\beta(c_i)\subseteq\mathcal{C}$ is the set of nodes that supply input to $c_i$\footnote{Note that this circuit representation of a computation can be provided by the programming language.}. Furthermore, let $\gamma(c_i)\in\mathcal{S}$ be the secret sharing scheme assigned to $c_i$. Then the monetary cost of executing a node $c_i\in\mathcal{C}$ is simply:
\begin{align}
	\mathtt{cost}(c_i) =  \mathtt{P}(\alpha(c_i), \gamma(c_i)) + \mathtt{N}(\alpha(c_i), \gamma(c_i)) +  \sum\limits_{c_j\in\beta(c_i)}{\mathtt{CP}(\gamma(c_j), \gamma(c_i)) + \mathtt{CN}(\gamma(c_j), \gamma(c_i))}
\end{align}

Using the above definitions, the optimal partitioning problem that we investigate in this paper can be formally defined as follows:

\begin{definition}
	Given the processing and network transfer cost for a set of operations $\mathcal{O}$ using a set of secret sharing schemes $\mathcal{S}$, the cost of conversion between different secret sharing schemes, and a circuit $\mathcal{C} = \{c_1, \ldots, c_m\}$ of $m$ nodes, where each node $c_i$ is assigned an operation $\alpha(c_i) \in \mathcal{O}$, assign a secret sharing scheme to each node $c_i$ such that the total monetary cost of executing the circuit with the assigned secret sharing schemes is minimal.

\begin{align}
	\begin{split}
	 \text{Given}: 	& \mathcal{O}, \mathcal{S}, \mathcal{C}, 
		 		\mathtt{P}(o, s)~\text{and}~\mathtt{N}(o,s)~\forall o \in \mathcal{O}, \forall s \in \delta(o), \\
				& \mathtt{CP}(s_i, s_j)~\text{and}~\mathtt{CN}(s_i, s_j)~\forall s_i, s_j \in \mathcal{S}, 
				 \alpha(c_i)~\forall c_i \in \mathcal{C} \\
	\text{Minimize}: 	& \sum\limits_{c_i \in \mathcal{C}}{cost(c_i)} 
	\text{~Subject to}: \gamma(c_i) \in \delta(\alpha(c_i)) ~ \forall c_i \in \mathcal{C}
	\end{split}
\end{align}
\end{definition}

The partitioning problem is NP-Hard, and the reduction can be simply done from the Integer Programming problem.

\subsection{Assumptions}
\label{sec:problem:assumptions}

We assume that the protocol between the two parties (i.e., the server and the client) is represented as a DAG tree. This can easily be assured by the secure multi-party implementation interface, or by some interpreter that allows programmers to implement the protocol using a custom programming language. The protocols implemented by that interface (or custom programming language) can be converted to the circuit representation. A similar assumption is also made by Kerschbaum et al., where the protocol written in custom language is transformed into a set of static operations~\cite{cite:kerschbaum2014automatic}. 


Finally, we assume that overall secure circuit evaluation is performed sequentially. In order to decrease the running time and reduce the processing cost, one may execute the operations specified by the non-dependent nodes in the circuit in parallel. Under such circumstances, our monetary cost model gives an upper bound on the expected monetary cost. Nevertheless, the cost model can be augmented to reflect this condition by considering the number of concurrent threads. However, the processing cost of executing operations under a secret sharing scheme would not decrease linearly as more threads are used, since some threads may overlap and block each other from uninterrupted execution. Due to such complications, the unit cost calculations may not be accurate.  For this reason, we will focus on the sequential protocol execution scenario in this work; and leave the multi-threaded implementations as a future work.

\vspace{-0.3cm}
\section{The Details of \sysname}
\label{sec:system}

\subsection{Architecture}
\label{sec:system:arch}

\sysname has three main parts: (i) The programming interface (API), (ii) the optimization module, and (iii) the SMC layer. Each \sysname part is responsible for a different task: Programming interface morphs user program into the circuit representation; the optimization module assigns secret sharing schemes to each node in the circuit using some heuristic; the SMC layer generates the executables using state-of-the-art cryptographic primitives and techniques. The user of our system is expected to insert two inputs:  specifications of the protocol that is going to be executed securely, and the unit costs for the operations and secret sharing schemes supported by the SMC framework. As we discuss later, we provide a benchmark suite to automatically learn these unit costs for any target cloud service to help the user. 

\textbf{User Inputs.} The user is assumed to know the secure protocol for the application that \sysname is used for. One input to our system is this protocol specification that can be either implemented using our C++ library or via a possible custom programming language, whose compiler turns the user input to an output compatible with our programming API. Next, the user has to input the unit monetary cost of the operations for the hardware specifications that secure executables will work on. We have implemented a set of benchmark applications that can be easily executed by a user to find the unit costs of each single operation. We would like to stress that this is a one-time operation per the tested cloud environment.  
 
\textbf{Programming API.} We implemented an extensible library in C++ programming language that allows a user of our system to implement a secure protocol (e.g., set intersection, biometric matching, etc.). We provided several operations in the API that will cover a variety of applications. Currently, the set of operations $\mathcal{O}$ include addition ($\mathtt{Add}$), subtraction ($\mathtt{Sub}$), multiplication ($\mathtt{Mul}$), greater ($\mathtt{Ge}$), equality ($\mathtt{Eq}$), multiplexer ($\mathtt{Mux}$), binary xor ($\mathtt{Xor}$), and binary and ($\mathtt{And}$). There are also two additional operations input ($\mathtt{In}$) and output ($\mathtt{Out}$) that allows the programmer to specify secret inputs to the circuit, and to learn the outputs of the protocol execution. 

As a proof of concept, we provided the interface as a C++ library that can be easily used to generate cost-optimal SMC executables. Additionally, this interface can be bound with the compiler of a custom programming language. Using this custom programming language, the user of our system can type the protocol specifications. Then the compiler can simply generate the C++ program that in turn uses \sysname programming interface. In any scenario, the programming interface morphs the protocol to the circuit representation discussed in Section~\ref{sec:problem}. 

\textbf{Optimizer.} Given the circuit representation of the user protocol and the secret sharing schemes that are provided by the SMC layer, the optimizer module applies one of the heuristics (cf. Section~\ref{sec:system:opt}) to assign secret sharing schemes to each node in the circuit. As we discussed previously, this module is responsible for finding the assignment that minimizes the monetary cost of executing the protocol securely. 

Due to the NP-Hard optimal partitioning problem, finding the optimal assignment may be impractical even for a slightly large circuit. Hence, we proposed several heuristics (cf. Section~\ref{sec:system:opt}) that tries to find a reasonable solution. In the background, \sysname applies each heuristic and chooses the one that gives the best result. 

\textbf{SMC Layer.} Once the secret sharing schemes are assigned to each node in the circuit, \sysname gives the circuit to the SMC layer to generate the SMC executables. There are several related works that provide mixed-protocol SMC tools, such as \emph{ABY} framework by Demmler et al~\cite{cite:demmler2015aby}, \emph{TASTY} framework by Henecka et al.~\cite{cite:henecka2010tasty}, \emph{Sharemind} framework by Bogdanov et al.~\cite{cite:bogdanov2008sharemind}, etc. We discuss all those tools in Section~\ref{sec:related}. The SMC layer in our \sysname benefits from such existing tools, and is responsible for automatically implementing the optimized circuit using the selected tool. We designed our system to enable easy integration with any possible SMC tool. Note that since the selected SMC tool provides the low-level implementation of the cryptographic primitives (e.g., oblivious transfer~\cite{cite:naor2001efficient, cite:rabin2005exchange}, multiplication triplets~\cite{cite:beaver1992efficient}, sharing and reconstructing secret inputs, etc.), we focus on optimizing user protocol using sharing assignments.

One clear connection between the SMC layer and the Optimizer module is the dependency on the selected SMC tool. Since each SMC tool may provide a different set of secret sharing mechanisms, the optimizer should perform the assignment such that the optimal circuit can actually be realized and executed by the SMC layer. For instance the \emph{ABY} framework provides three different secret sharing schemes: arithmetic sharing, boolean sharing, and garbled circuit sharing. On the other hand, the \emph{TASTY} framework allows additively homomorphic sharing and garbled circuit sharing. In any case, the user is responsible for running our sample applications only once for the tested environment and input the statistics data, while the SMC layer and the Optimizer exchanges information about the available secret sharing schemes based on selected SMC tool.

\subsection{Optimization Heuristics}
\label{sec:system:opt}
In our work, we developed two new heuristics to solve the optimal partitioning problem in addition to two existing heuristics. Below, we provide the details of these heuristics.

\textbf{Bottom-up Heuristic.} The key idea in this heuristic is to assign \emph{optimal} secret sharing scheme to the nodes in their topological order in the circuit. When a node $c_i \in \mathcal{C}$ is to be processed, the heuristic first assigns sharing schemes to the nodes that provide input to $c_i$. Based on the values assigned to the \emph{children}, this heuristic selects the scheme that minimizes the expected monetary cost for $c_i$. 

\textbf{Top-Down Heuristic.} In this technique, we process the nodes in the circuit in the opposite manner compared to the previous heuristic, and assign secret sharing schemes to the higher level nodes first and iterate down to the lower levels. The idea in this heuristic is to assign the scheme that minimizes the cost of the current node given that the schemes for the nodes that it is input to are already known. Assume that the secret sharing scheme for the node $c_k$ is set to $s_k$ previously. Now when assigning the secret sharing scheme for the node $c_i$, this heuristic takes into account that the result of the node $c_i$ should be converted to $s_k$. The \emph{optimal} decision is made with this consideration in mind. 

\textbf{Fixed Secret Sharing.} In this optimization heuristic, each node in the circuit is assigned the same secret sharing scheme. However, in some SMC tools, certain secret sharing schemes may not necessarily support each single operation (e.g., Arithmetic sharing in \emph{ABY} framework does not support $\mathtt{And}$, $\mathtt{Xor}$, and $\mathtt{Mux}$). In such a case, this heuristic selects one scheme that supports each \sysname operation. One common secret sharing scheme that is included in almost all SMC tools and supports each operation is garbled circuit sharing that implements Yao's garbled circuit protocol~\cite{cite:yao1982protocols}. Using this heuristic, we can measure the monetary cost of executing the user protocol by a single secret sharing scheme (e.g., pure SMC by garbled circuit sharing).

\textbf{Hill-Climbing.} This heuristic is based on the technique of Kerschbaum et al.~\cite{cite:kerschbaum2014automatic}. The basic idea is to start by assigning a common secret sharing scheme (e.g., garbled circuit sharing) to each node in the circuit. Next, we check if the total cost can be reduced by changing the current secret sharing scheme of a node. This loop continues till the total cost cannot be improved by any further assignment. 

\vspace{-0.6cm}
\begin{algorithm}[hbt]
\footnotesize
	\begin{algorithmic}[1]
		\State \textbf{Known:} $\mathcal{O}$, $\mathcal{S}$, $\mathcal{C}$, 
		\State \textbf{Known:} $\mathtt{P}(o, s)$~and~$\mathtt{N}(o,s)~\forall o \in \mathcal{O}, \forall s \in \delta(o)$, 
		\State \textbf{Known:} $\mathtt{CP}(s_i, s_j)$~and~$\mathtt{CN}(s_i, s_j) ~ \forall s_i, s_j \in \mathcal{S}$, 
		\State \textbf{Known:} $\alpha(c_i)~\forall c_i \in \mathcal{C}$

		\Function{OptimizeCircuit}{$\mathcal{C}$, $s_c \in \mathcal{S}$}
		\ForAll{$c_i \in \mathcal{C}$} 
			\State $\gamma(c_i) \gets s_c$ 
		\EndFor

		\Repeat
			\State $hasChanged \gets 0$
			\ForAll{$c_i \in \mathcal{C}$}
				\State $s \gets \argmin\limits_{s_j \in \delta(\alpha(c_i))} cost(c_i)$
				\If{$s \neq \gamma(c_i)$}
					\State $\gamma(c_i) \gets s$
					\State $hasChanged \gets 1$
				\EndIf
			\EndFor
		\Until{$hasChanged = 1$}
		\EndFunction
	\end{algorithmic}
	
	\caption{Hill-Climbing heuristic that assigns secret sharing schemes to the nodes in the circuit}
	\label{alg:heuristic:hill}
\end{algorithm}

\vspace{-0.45cm}
Algorithm~\ref{alg:heuristic:hill} details the hill-climbing heuristic. It takes a circuit $\mathcal{C}$ and an initial secret sharing scheme $s_c \in \mathcal{S}$ that supports each operation in the circuit as inputs (step 5). First, this given scheme is assigned to each node in the circuit (steps 6-8). Then, for each node in the circuit, it checks if the total cost of this node can be reduced by changing its current secret sharing scheme to another one(step 12). If successful, the change is processed and a flag is set to true to signal another round of loop iteration (steps 13-16). This iteration continue until no improvement can be made (step 18).

\subsection{Prototype Implementation}
\label{sec:system:proto}

As a proof of concept, we implemented a prototype of \sysname using C++ programming language. As discussed before, using the library of  our programming interface, a user can implement protocols that will be compiled into SMC executables. For the SMC layer, we used the ABY framework by Demmler et al.~\cite{cite:demmler2015aby}. The set of sharing schemes that are provided by \emph{ABY} are Arithmetic, Boolean, and garbled circuit sharing (that we refer to as the Yao sharing). Only two operations (i.e., $\mathtt{Add}$ and $\mathtt{Mul}$) are supported for the Arithmetic sharing, which are implemented using multiplications triplets by Beaver et al.~\cite{cite:beaver1992efficient}. Boolean sharing support each \sysname operation, while the low-level implementations by ABY are performed using Goldreich-Micali-Wigderson (GMW) protocol~\cite{cite:goldwasser1987play}. Finally, Yao sharing support each operation in \sysname and executes the operations using Yao's garbled circuit protocol~\cite{cite:yao1982protocols}. Further information about the \emph{ABY} framework can be found in ~\cite{cite:demmler2015aby}.

\vspace{-0.5cm}
\section{Cost Profiling}
\label{sec:cost}

\subsection{Experiment Setup}
\label{sec:cost:setup}

\textbf{Virtual Machines.} We used four different virtual machine models in Amazon EC2 cloud service. Table~\ref{table:cost:ec2_types} shows an overview of the VM specifications as of May 2015. We focused on selecting two different VM \emph{types}. The first two, m3.medium and m3.large, are \emph{memory optimized} models that provide better memory throughput. On the other hand, the last two VM models are \emph{compute optimized} that give faster processing throughput. As can be seen from table~\ref{table:cost:ec2_types}, the number of virtual central processing units (vCPU) and the memory sizes differ in the selected instances to test whether a faster (and more expensive) VM model provides any monetary gains.

\vspace{-0.4cm}
\begin{table}[hbt]
	\scriptsize
	\centering
	\begin{tabular}{|c|c|c|c|}

	\hline
	Model & CPU Type (Intel Xeon) &  vCPU & Memory (GB) \\
	\hline
	m3.medium	& E5-2670 v2 & 1 & 3.75 \\
	m3.large		& E5-2670 v2 & 2 & 7.5 \\
	c4.large		& E5-2666 v3 & 2 & 3.75 \\
	c4.xlarge		& E5-2666 v3 & 4 & 7.5 \\
	\hline
	\end{tabular}

	\caption{The specifications of four different Amazon EC2 VM models as of May 2015.}
	\label{table:cost:ec2_types}
\end{table}

\vspace{-0.9cm}
\textbf{Unit Costs.} We selected two Amazon EC2 regions to initiate our VMs: North Virginia and Tokyo. As of May 2015, customers of Amazon EC2 service are charged hourly, whereas the unit cost of running a VM varies with respect to its specifications. Table~\ref{table:cost:ec2_cost} gives the unit costs of running a VM for an hour in North Virginia or Tokyo Amazon EC2 regions. The network transfer cost also varies with respect to the Amazon EC2 region. The unit network transfer cost in North Virginia for incoming and outgoing network is $\$0.01/GB$ and $\$0.02/GB$, respectively. For the Tokyo EC2 region, the network costs are $\$0.01/GB$ for incoming and $\$0.09/GB$ for outgoing traffic.

\vspace{-0.3cm}
\begin{table}[hbt]
	\scriptsize
	\centering
	\begin{tabular}{|c|c|c|}
	\cline{2-3}
	\multicolumn{1}{c}{} & \multicolumn{2}{|c|}{Unit Cost ($\cent/hour$)} \\
	\cline{2-3}
	\multicolumn{1}{c|}{} & N. Virginia & Tokyo \\
	\hline
	m3.medium	& 7		& 10.1 \\
	m3.large		& 14		& 20.3 \\
	c4.large		& 11.6	& 14.7 \\	
	c4.xlarge		& 23.2	& 29.4 \\
	\hline
	\end{tabular}

	\caption{The average cost of running an Amazon EC2 VM model for an hour in two different AWS EC2 regions.}
	\label{table:cost:ec2_cost}
\end{table}

\vspace{-1.0cm}
\textbf{Scenarios.} We consider and perform cost measurements for two different scenarios. In the first scenario that we refer to as the \emph{Intra-Region} scenario, two parties in the computation initiate their VMs in the same Amazon EC2 region, North Virginia. Since Amazon does not charge network transfer between two VMs in the same region, the network cost for each gate is simply zero. On the other hand, the computation cost is simply running two VMs (one for the server, one for the client) for a given specification. 

In the second scenario that we refer to as the \emph{Inter-Region} scenario, one VM is located in North Virginia, while the other is in Tokyo. The processing cost is the cost of running two VMs in two different regions at the same time. We take the unit network transmission cost of transferring as $\$0.065/GB$. The reason is that there can be two cases: the server is in Tokyo and the client is in North Virginia, or vice versa. 
Since the server and the client roles can interchange, we take the unit cost of two cases.

Note that the server and the client uses the same Amazon EC2 VM model (i.e., m3.medium, etc.) in a given scenario for our experiments. 

\textbf{Secret Sharing Types.} Since we used the \emph{ABY} framework as our SMC layer, the secret sharing types that are profiled in this section are the Arithmetic, Boolean, and Yao secret sharing. We refer the reader to the original work for more details on the secret sharing types~\cite{cite:demmler2015aby}.

\textbf{Miscellaneous.} In order to get the average cost of executing a single operation $o_i$, we generated circuits that have $1000$ sequential $o_i$ operations on $32$-bit inputs. We divide the total time by $1000$ to get the average execution time for $o_i$. We repeat each experiment $10$ times and multiply the resulting number with the unit processing and network transfer costs. 

\vspace{-0.5cm}
\subsection{Results}
\label{sec:cost:results}

Table~\ref{table:cost:local_computation} shows the average computational cost of executing an operation in \emph{Intra-Region} scenario. Note that the unit for each number in the table is $\cent10^{-10}$. We calculated results for four different Amazon EC2 models for each operation provided in \sysname, along with the secret sharing conversion operations that are transparent to the user. For each operation and model, the table shows the results of executing the operation with Arithmetic, Boolean, or Yao sharing in top-down order. Since the arithmetic sharing does not support operations other than $\mathtt{Add}$ and $\mathtt{Mul}$, the results are simply null for other operations. \textit{Moreover, the conversion costs are sharing independent and are the same for each sharing type.} 

We see that executing $\mathtt{Add}$ and $\mathtt{Mul}$ using Arithmetic sharing is cheaper than the other two in all VM models, while conversion from Yao and Boolean sharing to Arithmetic sharing differs with the VM model. For instance, in m3.medium and m3.large, it is cheaper to switch to Arithmetic sharing and execute an $\mathtt{Add}$ operation. On the other hand in c4.large and c4.xlarge, the conversion cost makes it more expensive to switch from Yao to Arithmetic and execute $\mathtt{Add}$, compared to using Yao sharing. For the $\mathtt{Mul}$ operation, Arithmetic sharing is favorable in all models. An interesting finding is the disadvantage of Boolean sharing. Other than $\mathtt{Add}$ and $\mathtt{Mul}$, in almost all cases, performing an operation is cheaper using Yao sharing. Performing $\mathtt{Xor}$ and comparison operations (i.e., $\mathtt{Eq}$ and $\mathtt{Ge}$) in Yao sharing is much cheaper than Boolean sharing.

\begin{table*}[hbt]
	\scriptsize
	\centering
	\begin{tabular}{|c|c|c|c|c|c|c|c|c|c|c|c|c|c|c|}
	
	\cline{2-15}
	 \multicolumn{1}{c|}{} & $\mathtt{Add}$ & $\mathtt{Sub}$ & $\mathtt{Mul}$ & $\mathtt{And}$ & $\mathtt{Xor}$ & $\mathtt{Mux}$ & $\mathtt{Eq}$ & $\mathtt{Ge}$ & $\mathtt{A2B}$ & $\mathtt{A2Y}$ & $\mathtt{B2A}$ & $\mathtt{B2Y}$ & $\mathtt{Y2A}$ & $\mathtt{Y2B}$ \\
	\hline

	\multirow{3}{*}{m3.medium}
		& 21.54 	& - 		& 472.69 	& - 		& - 		& - 		& - 		& - 									&  	&  	&  	&  	&  	&  \\
		& 402.96 	& 173.00 	& 2791.63 	& 412.06 	& 13.22 	& 414.20 	& 28.01 	& 399.73 	& 156.99 	& 143.41 	& 38.77 	& 51.64 	& 73.32 	& 35.93 \\
		& 82.11 	& 162.79 	& 3125.30 	& 40.08 	& 17.00 	& 56.48 	& 11.83 	& 12.83 	&  	&  	&  	&  	&  	&   \\
	\hline

	\multirow{3}{*}{m3.large}
		& 50.30	& -		& 781.58	& - 		& - 		& - 		& - 		& - 							&  	&  	&  	&  	&  	&  \\
		& 466.67	& 202.35	& 3658.97	& 783.3	& 8.03	& 792.08	& 51.60	& 746.67	& 167.11	& 155.86	& 44.37	& 65.44	& 120.87	& 46.11 \\
		& 93.46	& 168.53	& 3240.53	& 44.13	& 19.60	& 62.62	& 9.22	& 12.49				&  	&  	&  	&  	&  	&  \\
	\hline		

	\multirow{3}{*}{c4.large}
		& 39.77	& -		& 284.15	& -		& -		& -		& -		& -		&  	&  	&  	&  	&  	&  \\
		& 310.03	& 159.57	& 2397.78	& 293.64	& 23.27	& 302.41	& 53.83	& 272.40	& 144.04	& 144.33 	& 22.91	& 60.36	& 95.31	& 43.94 \\
		& 72.08	& 115.56	& 2215.21	& 42.90	& 31.23	& 52.39	& 26.01	& 21.28	&  	&  	&  	&  	&  	&  \\
	\hline

	\multirow{3}{*}{c4.xlarge}
		& 123.72	& -		& 623.84	& -		& -		& -		& -		& -											&  	&  	&  	&  	&  	&  \\
		& 661.61	& 375.56 	& 5534.74	& 649	& 62.25	& 664.42	& 145.65	& 602.45	& 321.06	& 295.51 	& 144.09	& 180.99	& 205.20	& 135.17 \\
		& 188.53	& 272.43 	& 4959	& 135.32	& 110.15	& 152.28	& 80.16	& 95.36	&  	&  	&  	&  	&  	&  \\
	\hline
	\end{tabular}

	\caption{The average computational cost of executing an operation in \emph{Intra-Region} scenario for four different Amazon EC2 VM models. The order of results from top to bottom in a given VM model is \emph{Arithmetic}, \emph{Boolean}, and \emph{Yao} sharing. The unit is $\cent10^{-10}$.}
	\label{table:cost:local_computation}
\end{table*}

Table~\ref{table:cost:online_computation} shows the average computation cost of executing an operation in \emph{Inter-Region} scenario. Note that the unit for each result in the table is $\cent10^{-6}$. One crucial observation is that the costs are almost $4$ fold higher in the \emph{Inter-Region} scenario with respect to the \emph{Intra-Region} scenario due to the communication cost induced by the long distance. It takes longer to exchange messages between two parties, thus resulting in longer execution times and more costs. By looking at the table, it is fair to say that Yao sharing outperform the other two sharing in many cases. Even performing a $\mathtt{Mul}$ operation is cheaper with Yao sharings with respect to Arithmetic sharings. $\mathtt{Add}$ is still the cheapest using Arithmetic sharing. However, the conversion cost from Boolean and Yao to Arithmetic may make it a costly choice. Except few cases, executing operations using Boolean sharing is too costly. 

\vspace{-0.3cm}
\begin{table*}[hbt]
	\scriptsize
	\centering
	\begin{tabular}{|c|c|c|c|c|c|c|c|c|c|c|c|c|c|c|}
	
	\cline{2-15}
	 \multicolumn{1}{c|}{} & $\mathtt{Add}$ & $\mathtt{Sub}$ & $\mathtt{Mul}$ & $\mathtt{And}$ & $\mathtt{Xor}$ & $\mathtt{Mux}$ & $\mathtt{Eq}$ & $\mathtt{Ge}$ & $\mathtt{A2B}$ & $\mathtt{A2Y}$ & $\mathtt{B2A}$ & $\mathtt{B2Y}$ & $\mathtt{Y2A}$ & $\mathtt{Y2B}$ \\
	\hline

	\multirow{3}{*}{m3.medium}
		& 2.90		& -		& 2134.72	& -		& -		& -		& -		& -		&  	&  	&  	&  	&  	&  \\
		& 219.54	& 137.15	& 3350.81	& 2246.32	& 2.86	& 2122.63	& 11.55	& 2233.19	& 28.35	& 28.12	& 18.99	& 24.07	& 25.39	& 14.56 \\
		& 14.18		& 18.74	& 339.26	& 15.09	& 5.80	& 15.24	& 7.12	& 6.80	&  	&  	&  	&  	&  	&  \\
	\hline

	\multirow{3}{*}{m3.large}
		& 5.57		& -		& 4132.46	& -		& -		& -		& -		& -		&  	&  	&  	&  	&  	&  \\
		& 415.67	& 227.24	& 5754.01	& 4155.78	& 4.45	& 4273.34	& 21.08	& 3934.18	& 54.45	& 59.73	& 34.41	& 42.93	& 43.87	& 23.47 \\
		& 21.12		& 32.49	& 387.23	& 20.05	& 7.87	& 25.17	& 11.97	& 11.83	&  	&  	&  	&  	&  	&  \\
	\hline		

	\multirow{3}{*}{c4.large}
		& 3.66		& -		& 2890.76	& -		& -		& -		& -		& -		&  	&  	&  	&  	&  	&  \\
		& 334.20	& 172.98	& 4123.40	& 2927.87	& 3.74	& 3182.07	& 16.36	& 2759.07	& 38.83	& 389.98	& 26.01	& 32.99	& 35.12	& 18.22 \\
		& 19.12		& 33.94	& 495.18	& 18.18	& 7.42	& 18.81	& 9.14	& 9.01	&  	&  	&  	&  	&  	&  \\
	\hline

	\multirow{3}{*}{c4.xlarge}
		& 7.49		& -		& 4954.67	& -		& -		& -		& -		& -		&  	&  	&  	&  	&  	&  \\
		& 691.22	& 349.52	& 7133.87	& 5546.53	& 7.48	& 5898.06	& 32.84	& 5440.71	& 77.32	& 77.39	& 53.01	& 67.84	& 73.17	& 37.06 \\
		& 37.78		& 60.53	& 868.64	& 36.77	& 14.16	& 36.81	& 18.51	& 18.15	&  	&  	&  	&  	&  	&  \\
	\hline
	\end{tabular}

	\caption{The average computational cost of executing an operation in \emph{Inter-Region} scenario for four different Amazon EC2 VM models. The order of results from top to bottom in a given VM model is \emph{Arithmetic}, \emph{Boolean}, and \emph{Yao} sharing. The unit is $\cent10^{-6}$.}
	\label{table:cost:online_computation}
\end{table*}

\vspace{-0.9cm}
Table~\ref{table:cost:online_network} shows the average network transfer cost of executing an operation in \emph{Inter-Region} scenario. Note that the unit for the numbers in the table is $\cent10^{-6}$. The results are applicable to each VM model, since the network transfer cost is calculated based on the size of the transferred data. One major observation is the results for the Yao sharing. If we had not included network transfer cost in our optimization formalization, we would end up favoring Yao sharing for $\mathtt{Add}$ and $\mathtt{Mul}$ operations. However, we see that network cost of executing $\mathtt{Mul}$ in Yao sharing introduces huge network cost, making it less favorable compared to using Arithmetic sharing. The difference between the network transfer and processing cost for executing $\mathtt{Mul}$ using Yao sharing is due to the nature of the garbled circuit technique. The circuit (or the gate) should be garbled by the server and sent to the client. The execution performed by the client may take less time with respect to executing it with Boolean or Arithmetic sharing. However, the network transfer introduces another dimension that should be considered. 

\vspace{-0.3cm}
\begin{table*}[hbt]
	\scriptsize
	\centering
	\begin{tabular}{|c|c|c|c|c|c|c|c|c|c|c|c|c|c|c|}
	
	\cline{2-15}
	 \multicolumn{1}{c|}{} & $\mathtt{Add}$ & $\mathtt{Sub}$ & $\mathtt{Mul}$ & $\mathtt{And}$ & $\mathtt{Xor}$ & $\mathtt{Mux}$ & $\mathtt{Eq}$ & $\mathtt{Ge}$ & $\mathtt{A2B}$ & $\mathtt{A2Y}$ & $\mathtt{B2A}$ & $\mathtt{B2Y}$ & $\mathtt{Y2A}$ & $\mathtt{Y2B}$ \\
	\hline
	\multirow{3}{*}{all VMs}
		& 0		&  -	& 75.14	& -		& -	& -		& -		& -		&  	&  	&  	&  	&  	&  \\
		& 490.1	& 0	& 4258.8	& 67.6	& 0	& 2.6		& 65.52	& 188.045	& 199.94	& 199.94	& 37.57	& 66.82	& 137.41	& 99.84 \\
		& 99.84	& 0	& 6289.92	& 99.84	& 0	& 99.84	& 96.72	& 99.84	&  	&  	&  	&  	&  	&  \\
	\hline
	\end{tabular}

	\caption{The average network cost of executing an operation in \emph{Inter-Region} scenario for all Amazon EC2 VM models. The order of results from top to bottom is \emph{Arithmetic}, \emph{Boolean}, and \emph{Yao} sharing. The unit is $\cent10^{-6}$.}
	\label{table:cost:online_network}
\end{table*}

\section{Case Studies}
\label{sec:case}

\subsection{Experiment Setup}
\label{sec:case:setup}

The setup for the case studies is exactly the same as described in Section~\ref{sec:cost:setup}. We perform tests in two scenarios, \emph{Intra-Region} and \emph{Inter-Region}, for four different Amazon EC2 VM models. We tested each scenario and VM model with four techniques: three optimization heuristics (i.e., \emph{Top-down}, \emph{Bottom-up}, and \emph{Hill Climbing}) and \emph{Pure-Yao}, which assigns Yao sharing (i.e., garbled circuit) to each node in the circuit. 

In addition to the monetary cost of running \sysname, we also measure and show the average running time of four techniques. Although our primary objective is to minimize the cost, we want to see if better performance is achieved by our heuristics. As mentioned in Section~\ref{sec:system:opt}., our optimization problem can be enhanced by introducing performance constraints ( e.g., the expected running time should be less than some threshold $t$). Given such a performance constraint, the heuristic solver may prune any solution that does not satisfy the estimated performance constraint. 

\subsection{Biometric Matching}
\label{sec:case:set}

\emph{Biometric matching} applications cover the two-party scenario, where (i) the server has a set of private entries, and (ii) a client holding its private entry wants to learn the closest entry in the server's dataset based on some similarity measure. There are various problems related to this case study (e.g., biometric identification~\cite{cite:evans2011efficient}, fingercode authentication~\cite{cite:barni2010privacy}, face recognition~\cite{cite:erkin2009privacy, cite:sadeghi2010efficient}, etc.). One of the commonly used distance metric is squared Euclidean distance. In this protocol, the server and the client go over the server's dataset one by one, which results by the client learning the entry with the minimal distance to its private input. We implemented this case study using our Programming API for a dataset of $30$ rows with $5$ attributes of $32$-bit numbers. 

Table~\ref{table:case:bio:run} shows the average running time for the \emph{Biometric Matching} case study, two different scenarios, four VM models, and four secret-sharing assignment heuristics. As expected, the performance is much better in the Intra-Region scenario, where the parties are in the same Amazon EC2 region. In all cases, applying any of the heuristics gives lower running times compared to the Pure-GC assignment (i.e., each node is assigned the Yao sharing). Moreover, our heuristic of Top-Down assignment gives the best running time for all VM models. It is $15\%$ better than the Hill Climbing heuristic of Kerschbaum et al~\cite{cite:kerschbaum2014automatic} in terms performance. For the Inter-Region scenario, we see that Pure-GC performs much better than the other techniques in terms of performance, except the last model c4.xlarge. Since the physical distance between the two parties is large (i.e., between Tokyo and North Virginia), network latency plays a vital role in the overall performance. And it is shown many times that Yao sharing (i.e., garbled circuit) is much better than any other solution in high-latency networks. Since our primary optimization objective is to minimize cost, not to minimize performance, the Inter-Region results are not surprising. 

\begin{table}[hbt]
	\scriptsize
	\centering
	\begin{tabular}{|c|c|c|c|c|c|}
	
	\cline{3-6}
	\multicolumn{2}{c|}{} & \multicolumn{4}{|c|}{Execution time ($ms$)} \\
	\cline{3-6}
	\multicolumn{2}{c|}{} & Pure-GC & Hill & TD & BU \\
	\hline
	
	\parbox[t]{2mm}{\multirow{4}{*}{\rotatebox[origin=c]{90}{INTRA}}} 
 & m3.med & 1229.481 & 398.02 & \textbf{348.43} & 416.944 \\
 & m3.large & 715.5147 & 355.2913 & \textbf{321.388} & 372.529 \\
 & c4.large & 577.333 & 293.544 & \textbf{256.363} & 284.378 \\
 & c4.xlarge & 561.602 & 352.197 & \textbf{292.522} & 326.393 \\		
	\hline		
	\parbox[t]{2mm}{\multirow{4}{*}{\rotatebox[origin=c]{90}{INTER}}} 
 & m3.med & \textbf{5518.04} & 6738.14 & 6237.37 & 6749.467 \\
 & m3.large & \textbf{4801.94} & 6096.73 & 6103.647 & 6083.193 \\
 & c4.large & \textbf{8731.577} & 35580.13 & 33267.77 & 35569.97 \\
 & c4.xlarge & 9118.693 & 6916.357 & \textbf{6381.093} & 6826.727 \\
	\hline
	\end{tabular}

	\caption{The average execution time for the \emph{Biometric Matching} case study in Amazon EC2 Cloud. The results are for two different scenarios, four different VM models, and four different techniques.}
	\label{table:case:bio:run}
\end{table}

\begin{table*}[hbt]
	\scriptsize
	\centering
	\begin{tabular}{|c|c|c|c|c|c|c|c|c|c|c|c|c|c|}
	
	\cline{3-14}
	\multicolumn{2}{c|}{} & \multicolumn{4}{|c|}{Computation Cost ($\cent10^{-3}$)} & \multicolumn{4}{|c|}{Network Cost ($\cent10^{-3}$)}& \multicolumn{4}{|c|}{Total Cost ($\cent10^{-3}$)} \\
	\cline{3-14}
	\multicolumn{2}{c|}{} & Pure-GC & Hill & TD & BU & Pure-GC & Hill & TD & BU & Pure-GC & Hill & TD & BU \\
	\hline
	
	\parbox[t]{2mm}{\multirow{4}{*}{\rotatebox[origin=c]{90}{INTRA}}} 
 & m3.med 		& 4.78 & 1.55 & \textbf{1.36} & 1.62 & 0.00 & 0.00 & 0.00 & 0.00 & 4.78 & 1.55 & \textbf{1.36} & 1.62 \\
 & m3.large 	& 5.57 & 2.76 & \textbf{2.50} & 2.90 & 0.00 & 0.00 & 0.00 & 0.00 & 5.57 & 2.76 & \textbf{2.50} & 2.90 \\
 & c4.large 	& 3.72 & 1.89 & \textbf{1.65} & 1.83 & 0.00 & 0.00 & 0.00 & 0.00 & 3.72 & 1.89 & \textbf{1.65} & 1.83 \\
 & c4.xlarge 	& 7.24 & 4.54 & \textbf{3.77} & 4.21 & 0.00 & 0.00 & 0.00 & 0.00 & 7.24 & 4.54 & \textbf{3.77} & 4.21 \\		
	\hline		
	\parbox[t]{2mm}{\multirow{4}{*}{\rotatebox[origin=c]{90}{INTER}}} 
 & m3.med 		& \textbf{27.74} 	& 33.88 	& 31.36 	& 33.93 	& 990.71 & 189.71 & \textbf{129.71} & 189.71 & 1018.46 & 223.58 & \textbf{161.07} & 223.64 \\
 & m3.large 	& \textbf{45.75} 	& 58.09 	& 58.15 	& 57.96 	& 990.71 & 189.71 & \textbf{129.71} & 189.71 & 1036.46 & 247.80 & \textbf{187.86} & 247.67 \\ 
 & c4.large 	& \textbf{63.79} 	& 259.93 	& 243.04 	& 259.86 	& 990.71 & 218.29 & \textbf{173.27} & 218.29 & 1054.50 & 478.23 & \textbf{416.31} & 478.15 \\
 & c4.xlarge 	& 133.23 	& 101.06 	& \textbf{93.23} 	& 99.75 	& 990.71 & 189.71 & \textbf{129.71} & 189.71 & 1123.95 & 290.76 & \textbf{222.94} & 289.45 \\		
	\hline
	\end{tabular}

	\caption{The average computational, network, and total cost of running the \emph{Biometric Matching} case study in Amazon EC2 Cloud. The results are for two different scenarios, four different VM models, and four different techniques.}
	\label{table:case:bio:cost}
\end{table*}

Table~\ref{table:case:bio:cost} shows the monetary cost for the \emph{Biometric Matching} case study with the aforementioned setup. In the Intra-Region scenario, we see that our Top-Down heuristic performs better than any other technique in all VM models. The reason is due to better assignment of secret sharing schemes to the nodes in the circuit for this particular case study compared to the other techniques. Note that the network communication cost within the same region is $0$ in Amazon EC2, which is why the network cost in Table~\ref{table:case:bio:cost} is simply $0$. In the Inter-Region scenario, Top-Down heuristic once again gives the cheapest assignments for all VM models. It performs $30\%$ better than the Hill Climbing heuristic. In terms of computation cost, we see that Pure-GC performs better due to the reasons discussed before (i.e., high network latency). However, in terms of total cost, Top-Down heuristic introduces up to $80\%$ reduction.

\vspace{-0.45cm}
\subsection{Matrix Multiplication}
\label{sec:case:matrix}

\emph{Matrix Multiplication} is a common building block of various applications (e.g., linear transformations, group theory, etc.). Given two matrices $A$ and $B$ of size $n$ by $n$, a common algorithm is the $O(n^3)$ algorithm that multiplies each row of $A$ with each column of $B$. We implemented this algorithm using our Programming API for two matrices of sizes $5x5$. 

\begin{table}[hbt]
	\scriptsize
	\centering
	\begin{tabular}{|c|c|c|c|c|c|}
	
	\cline{3-6}
	\multicolumn{2}{c|}{} & \multicolumn{4}{|c|}{Execution time ($ms$)} \\
	\cline{3-6}
	\multicolumn{2}{c|}{} & Pure-GC & Hill & TD & BU \\
	\hline
	
	\parbox[t]{2mm}{\multirow{4}{*}{\rotatebox[origin=c]{90}{INTRA}}} 
 & m3.med & 939.70 & 104.54 & 132.94 & \textbf{55.61} \\
 & m3.large & 504.76 & 18.00 & 16.64 & \textbf{16.30} \\
 & c4.large & 385.30 & \textbf{84.88} & 85.85 & \textbf{84.88} \\
 & c4.xlarge & 388.16 & 82.14 & \textbf{80.09} & 84.52 \\		
	\hline		
	\parbox[t]{2mm}{\multirow{4}{*}{\rotatebox[origin=c]{90}{INTER}}} 
 & m3.med & 5932.34 & 2258.09 & 2237.64 & \textbf{2211.83} \\
 & m3.large & 4175.28 & 2243.93 & \textbf{2234.97} & 2240.41 \\
 & c4.large & 8777.51 & \textbf{2247.47} & 2248.17 & 2253.02 \\
 & c4.xlarge & 7079.163 & 2273.117 & \textbf{2246.723} & 2257.253 \\
	\hline
	\end{tabular}

	\caption{The average execution time for the \emph{Matrix Multiplication} case study in Amazon EC2 Cloud. The results are for two different scenarios, four different VM models, and four different techniques.}
	\label{table:case:matrix:run}
\end{table}

\vspace{-0.4cm}
Table~\ref{table:case:matrix:run} shows the average running time for the \emph{Matrix Multiplication} case study for two different scenarios, four VM models, and four secret-sharing assignment heuristics. In the Intra-Region scenario, our heuristic of Bottom-Up gives better performance for the first $3$ VM models, while the Hill Climbing technique is the fastest for the c4.xlarge model. Once again, we focus on minimizing the monetary cost, not the running time. Still, our techniques are better than the existing technique of Kerschbaum et al. in all but one case. A similar statement is also true for the Inter-Region scenario: Our techniques perform better in $3$ different VM models, while Hill Climbing is better in c4.large model. Compared to the Pure-GC technique, all heuristics perform much better in terms of performance, reducing the running time up to $60\%$ in most cases. 

\vspace{-0.5cm}
\begin{table*}[hbt]
	\scriptsize
	\centering
	\begin{tabular}{|c|c|c|c|c|c|c|c|c|c|c|c|c|c|}
	
	\cline{3-14}
	\multicolumn{2}{c|}{} & \multicolumn{4}{|c|}{Computation Cost ($\cent10^{-3}$)} & \multicolumn{4}{|c|}{Network Cost ($\cent10^{-3}$)}& \multicolumn{4}{|c|}{Total Cost ($\cent10^{-3}$)} \\
	\cline{3-14}
	\multicolumn{2}{c|}{} & Pure-GC & Hill & TD & BU & Pure-GC & Hill & TD & BU & Pure-GC & Hill & TD & BU \\
	\hline
	
	\parbox[t]{2mm}{\multirow{4}{*}{\rotatebox[origin=c]{90}{INTRA}}} 
 & m3.med & 3.65 & 0.41 & 0.52 & \textbf{0.22} & 0.00 & 0.00 & 0.00 & 0.00 & 3.65 & 0.41 & 0.52 & \textbf{0.22} \\
 & m3.large & 3.93 & 0.14 & \textbf{0.13} & \textbf{0.13} & 0.00 & 0.00 & 0.00 & 0.00 & 3.93 & 0.14 & \textbf{0.13} & \textbf{0.13} \\
 & c4.large & 2.48 & \textbf{0.55} & \textbf{0.55} & \textbf{0.55} & 0.00 & 0.00 & 0.00 & 0.00 & 2.48 & \textbf{0.55} & \textbf{0.55} & \textbf{0.55} \\
 & c4.xlarge & 5.00 & 1.06 & \textbf{1.03} & 1.09 & 0.00 & 0.00 & 0.00 & 0.00 & 5.00 & 1.06 & \textbf{1.03} & 1.09 \\		
	\hline		
	\parbox[t]{2mm}{\multirow{4}{*}{\rotatebox[origin=c]{90}{INTER}}} 
 & m3.med & 29.83 & 11.35 & 11.25 & \textbf{11.12} & 796.22 & \textbf{9.39} & \textbf{9.39} & \textbf{9.39} & 826.05 & 20.75 & 20.64 & \textbf{20.51} \\
 & m3.large & 39.78 & 21.38 & \textbf{21.29} & 21.35 & 796.22 & \textbf{9.39} & \textbf{9.39} & \textbf{9.39} & 836.01 & 30.77 & \textbf{30.69} & 30.74 \\
 & c4.large & 64.12 & \textbf{16.42} & \textbf{16.42} & 16.46 & 796.22 & \textbf{9.39} & \textbf{9.39} & \textbf{9.39} & 860.35 & \textbf{25.81} & 25.82 & 25.85 \\ 
 & c4.xlarge & 103.43 & 33.21 & \textbf{32.83} & 32.98 & 796.22 & \textbf{9.39} & \textbf{9.39} & \textbf{9.39} & 899.66 & 42.61 & \textbf{42.22} & 42.37	\\	
	\hline
	\end{tabular}

	\caption{The average computational, network, and total cost of running the \emph{Matrix Multiplication} case study in Amazon EC2 Cloud. The results are for two different scenarios, four different VM models, and four different techniques.}
	\label{table:case:matrix:cost}
\end{table*}

\vspace{-1.0cm}
Table~\ref{table:case:matrix:cost} shows the monetary cost for the \emph{Matrix Multiplication} case study with the aforementioned setup. For the Intra-Region scenario, our techniques both perform better than or equal to the Hill Climbing technique in all VM models. Excluding the c4.xlarge model, our Bottom-Up technique wins the head to head comparison with other techniques. For the Inter-Region scenario, we see that except the c4.large VM model, our techniques give less total monetary cost than the other ones. In terms of the network cost, all three heuristics perform the same. However, the difference in the computation cost decides the leading technique in terms of total cost. Note that the differences between the heuristics is extremely small.

One final generic observation for both case studies is the importance of VM model selection. Based on our case study results, it is shown that choosing faster and more expensive VM model does not necessarily produce cheaper SMC executions. We see that for both case studies, in most cases, memory optimized VM types (i.e., m3.medium and m3.large) result in cheaper SMC executions, while executing the same user protocol in compute optimized VM types (i.e., c4.large and c4.xlarge) is a lot more expensive. 

\vspace{-0.4cm}
\section{Related Work}
\label{sec:related}

\vspace{-0.409cm}
There are many different works that abstract SMC implementation and provide users a user-friendly way to generate SMC executables. Holzer et al~\cite{cite:holzer2012secure} proposed a compiler that translates programs written in ANSI-C to their optimal garbled circuit executables. The authors modify the existing compiler tools, which already optimize the program in the background. The \emph{VIFF} framework provides a tradeoff between execution termination and efficiency, and approaches the SMC execution in an asynchronous manner~\cite{cite:damgaard2009asynchronous}. Multiple operations can be executed in parallel, further reducing the performance overhead, provided that the protocol is not guaranteed to terminate. 

\emph{Fairplay} and later extended to \emph{FairplanMP} were among the first efforts to introduce SMC executable generations in the background~\cite{cite:malkhi2004fairplay, cite:ben2008fairplaymp}. \emph{Fairplay} allows a user to implement its protocol in \emph{SFDL}, a custom intermediate language. Then, the compiler transforms the user protocol into garbled circuit executables, which can be executed by two parties in \emph{Fairplay}, and by multiple parties in \emph{FairplayMP}. \emph{GraphSC} by Nayak et al. provide a similar intermediate language that is specifically designed for graph-based algorithms~\cite{cite:nayak2015graphsc}. It aims to minimize the performance overhead by greatly paralleling the computations that are represented as garbled circuits. The work of Mood et al. provide a library to generate garbled circuit executables for mobile applications~\cite{cite:mood2012memory}. \emph{ObliVM} similarly translates programs written in custom language to their garbled circuit representations~\cite{cite:wang2015oblivm}. \emph{TinyGarble} performs further optimizations on the garbled circuit generation~\cite{cite:songhori2015tinygarble}. Finally, \emph{Wysteria} is a strongly typed intermediate language that once again generate garbled circuit executables~\cite{cite:rastogi2014wysteria}. None of the mentioned works can be used in our framework, since they provide only a single secret sharing scheme. Further improvements on the performance, and thus monetary cost, can not be achieved by switching between multiple schemes. 

One SMC tool that is suitable to (and also used in) our framework in SMC layer is the \emph{ABY} framework~\cite{cite:demmler2015aby}. It makes use of state-of-the-art optimization for each single operation, and provides a C++ library to implement user protocols. The secret sharing schemes that are supported in \emph{ABY} are Arithmetic, Boolean, and Yao's garbled circuit. The \emph{L1} framework is another SMC framework, written in Java, and supports the same set of secret sharing schemes~\cite{cite:schropfer2011l1}. \emph{Sharemind} provides additive secret sharing and Yao's garbled circuit as the available secret sharing schemes, and allows a user to generate SMC executables of its protocol~\cite{cite:bogdanov2008sharemind}. The work of Choudhury et al. provide a framework that uses garbled circuit and fully homomorphic encryption as the secret sharing schemes~\cite{cite:choudhury2013between}. \emph{Tasty} allows additively homomorphic encryption and garbled circuits as the secret sharing schemes~\cite{cite:henecka2010tasty}. None of the mentioned works consider optimizing the performance (or monetary cost) by selecting different secret sharing schemes for different statements (or nodes in the circuit). 

One work that is closely related to ours is the work of Kerschbaum et al.~\cite{cite:kerschbaum2014automatic}. However, their work differs from ours in terms of the focus of optimization: We aim to minimize the monetary cost of executing the user protocol in SMC, while they solely focus on the performance. As discussed in Section~\ref{sec:cost:results}, introducing additional dimensions such as network transmission cost may vary the secret sharing selection for a node during the optimization process.

\vspace{-0.4cm}
\section{Conclusion}
\label{sec:conclusion}

In this work, we propose \sysname, an SMC framework that aims to minimize the monetary cost of executing SMC protocols in the cloud. We performed extensive cost profiling for the Amazon EC2 cloud service. We tested four different VM models and two scenarios (i.e., Inter-Region and Intra-Region). Moreover, we leveraged the gathered statistics and applied our system  two case studies: Biometric matching and matrix multiplication. We showed that the cost of executing SMC using our heuristics is up to $96\%$ and $30\%$ less than using pure garbled circuit and Hill-Climbing, respectively. Moreover, we conclude that purchasing faster and more expensive VM model from Amazon EC2 does not necessarily reduce the total monetary cost of executing SMC protocols. In general, compute optimized VMs result in more expenses, while memory optimized ones produce cheaper SMC executions.

\bibliographystyle{abbrv}
\bibliography{references}

\begin{thebibliography}{10}

\bibitem{cite:barni2010privacy}
M.~Barni, T.~Bianchi, D.~Catalano, M.~Di~Raimondo, R.~Donida~Labati, P.~Failla,
  D.~Fiore, R.~Lazzeretti, V.~Piuri, F.~Scotti, et~al.
\newblock Privacy-preserving fingercode authentication.
\newblock In {\em ACM workshop on Multimedia and security}, pages 231--240.
  ACM, 2010.

\bibitem{cite:beaver1992efficient}
D.~Beaver.
\newblock Efficient multiparty protocols using circuit randomization.
\newblock In {\em Advances in Cryptology}, pages 420--432. Springer, 1992.

\bibitem{cite:ben2008fairplaymp}
A.~Ben-David, N.~Nisan, and B.~Pinkas.
\newblock Fairplaymp: a system for secure multi-party computation.
\newblock In {\em ACM CCS}, pages 257--266, 2008.

\bibitem{cite:bogdanov2008sharemind}
D.~Bogdanov, S.~Laur, and J.~Willemson.
\newblock Sharemind: A framework for fast privacy-preserving computations.
\newblock In {\em ESORICS}, pages 192--206. Springer, 2008.

\bibitem{cite:choudhury2013between}
A.~Choudhury, J.~Loftus, E.~Orsini, A.~Patra, and N.~P. Smart.
\newblock Between a rock and a hard place: Interpolating between mpc and fhe.
\newblock In {\em ASIACRYPT}, pages 221--240. Springer, 2013.

\bibitem{cite:cramer2000general}
R.~Cramer, I.~Damg{\aa}rd, and U.~Maurer.
\newblock General secure multi-party computation from any linear secret-sharing
  scheme.
\newblock In {\em EUROCRYPT 2000}, pages 316--334. Springer, 2000.

\bibitem{cite:damgaard2009asynchronous}
I.~Damg{\aa}rd, M.~Geisler, M.~Kr{\o}igaard, and J.~B. Nielsen.
\newblock Asynchronous multiparty computation: Theory and implementation.
\newblock In {\em PKC}, pages 160--179. Springer, 2009.

\bibitem{cite:demmler2015aby}
D.~Demmler, T.~Schneider, and M.~Zohner.
\newblock {ABY} - {A} framework for efficient mixed-protocol secure two-party
  computation.
\newblock In {\em NDSS}, 2015.

\bibitem{cite:erkin2009privacy}
Z.~Erkin, M.~Franz, J.~Guajardo, S.~Katzenbeisser, I.~Lagendijk, and T.~Toft.
\newblock Privacy-preserving face recognition.
\newblock In {\em PETS}, pages 235--253. Springer, 2009.

\bibitem{cite:evans2011efficient}
D.~Evans, Y.~Huang, J.~Katz, and L.~Malka.
\newblock Efficient privacy-preserving biometric identification.
\newblock In {\em NDSS}, 2011.

\bibitem{cite:forbes}
Forbes.
\newblock Cloud computing: United states businesses will spend \$13 billion on
  it.
\newblock
  http://www.forbes.com/sites/tjmccue/2014/01/29/cloud-computing-united-states-businesses-will-spend-13-billion-on-it,
  2014.

\bibitem{cite:goldwasser1987play}
S.~Goldwasser, S.~Micali, and A.~Wigderson.
\newblock How to play any mental game, or a completeness theorem for protocols
  with an honest majority.
\newblock In {\em ACM STOC}, volume~87, pages 218--229, 1987.

\bibitem{cite:henecka2010tasty}
W.~Henecka, A.-R. Sadeghi, T.~Schneider, I.~Wehrenberg, et~al.
\newblock Tasty: tool for automating secure two-party computations.
\newblock In {\em ACM CCS}, pages 451--462, 2010.

\bibitem{cite:holzer2012secure}
A.~Holzer, M.~Franz, S.~Katzenbeisser, and H.~Veith.
\newblock Secure two-party computations in ansi c.
\newblock In {\em ACM CCS}, pages 772--783, 2012.

\bibitem{cite:kerschbaum2014automatic}
F.~Kerschbaum, T.~Schneider, and A.~Schr{\"o}pfer.
\newblock Automatic protocol selection in secure two-party computations.
\newblock In {\em ACNS}, pages 566--584. Springer, 2014.

\bibitem{cite:malkhi2004fairplay}
D.~Malkhi, N.~Nisan, B.~Pinkas, Y.~Sella, et~al.
\newblock Fairplay-secure two-party computation system.
\newblock In {\em USENIX Securit}, volume~4, 2004.

\bibitem{cite:mood2012memory}
B.~Mood, L.~Letaw, and K.~Butler.
\newblock Memory-efficient garbled circuit generation for mobile devices.
\newblock In {\em Financial Cryptography and Data Security}, pages 254--268.
  Springer, 2012.

\bibitem{cite:naor2001efficient}
M.~Naor and B.~Pinkas.
\newblock Efficient oblivious transfer protocols.
\newblock In {\em SIAM}, pages 448--457, 2001.

\bibitem{cite:nayak2015graphsc}
K.~Nayak, X.~S. Wang, S.~Ioannidis, U.~Weinsberg, N.~Taft, and E.~Shi.
\newblock Graphsc: Parallel secure computation made easy.
\newblock In {\em IEEE S \& P}, 2015.

\bibitem{cite:rabin2005exchange}
M.~O. Rabin.
\newblock How to exchange secrets with oblivious transfer.
\newblock {\em IACR Cryptology ePrint Archive}, 2005:187, 2005.

\bibitem{cite:rastogi2014wysteria}
A.~Rastogi, M.~A. Hammer, and M.~Hicks.
\newblock Wysteria: A programming language for generic, mixed-mode multiparty
  computations.
\newblock In {\em IEEE S \& P}, pages 655--670, 2014.

\bibitem{cite:sadeghi2010efficient}
A.-R. Sadeghi, T.~Schneider, and I.~Wehrenberg.
\newblock Efficient privacy-preserving face recognition.
\newblock In {\em ICISC}, pages 229--244. Springer, 2010.

\bibitem{cite:schropfer2011l1}
A.~Schropfer, F.~Kerschbaum, and G.~Muller.
\newblock L1-an intermediate language for mixed-protocol secure computation.
\newblock In {\em IEEE COMPSAC}, pages 298--307, 2011.

\bibitem{cite:shamir1979share}
A.~Shamir.
\newblock How to share a secret.
\newblock {\em Communications of the ACM}, 22(11):612--613, 1979.

\bibitem{cite:songhori2015tinygarble}
E.~M. Songhori, S.~U. Hussain, A.-R. Sadeghi, T.~Schneider, and F.~Koushanfar.
\newblock Tinygarble: Highly compressed and scalable sequential garbled
  circuits.
\newblock In {\em IEEE S \& P}, 2015.

\bibitem{cite:wang2015oblivm}
X.~Wang, C.~Liu, K.~Nayak, Y.~Huang, and E.~Shi.
\newblock Oblivm: A programming framework for secure computation.
\newblock In {\em IEEE S \& P}, 2015.

\bibitem{cite:yao1982protocols}
A.~C. Yao.
\newblock Protocols for secure computations.
\newblock In {\em IEEE ASFCS}, pages 160--164. IEEE, 1982.

\bibitem{cite:yao1986generate}
A.~C.-C. Yao.
\newblock How to generate and exchange secrets.
\newblock In {\em IEEE FOCS}, pages 162--167, 1986.

\end{thebibliography}

\end{document}